# Review: CPG as a controller for biomimetic floating robots

Zharinov A. I.[1,2] Tsybina Y. A.[2] Gordleeva S. Y. [1,2].

zharinov@neuro.nnov.ru
1. Immanuel Kant Baltic Federal University
2. Lobachevsky State University of Nizhni Novgorod

## Abstract

All vertebrates are capable of performing various types of physical activity. Locomotor patterns are created by the cyclical coordinated work of the skeletal muscles. The organization of such a system in living organisms is responsible for networks of interconnected populations of neurons capable of forming rhythmic activity - CPG. As biologists have established, CPG is a key mechanism causing adaptive and versatile locomotion in animals. Using a system based on biological principles as a controller for robotic devices will avoid many of the problems that are present in standard control loops. In particular, it becomes possible not only plausible modeling and study of the peculiarities of animal locomotion (in simple cases), but also smooth switching between different types of activity depending on environmental conditions, i.e. creation of autonomous adaptive systems. In this review, a description of the work of recent years is given, in which the CPG is used as a controller for various robotic animals. We focused on the views showing undulating movements, i.e. floating and amphibians. Each separate section deals with a certain kind. At the same time, the works are arranged in the chronological order of their publication. At the end of each section, there is a summary that categorizes the articles described by the type of CPG used.

## Introduction

All vertebrates are capable of performing various types of physical activity. Locomotor patterns are created by the cyclical coordinated work of the muscles of the skeletal muscles. A feature of such a control system is its ability to generate the necessary rhythms without control signals from the brain. The communities of neurons that regulate locomotor patterns are called Central Rhythm Generators (Ijspeert, Crespi. 2005; Lachat et al. 2006; Wang, Hu et al. 2013; Li et al. 2014a). As established by biologists, CPG is a key mechanism causing adaptive and versatile locomotion in animals (J. Ijspeert, 2008, J. Yu, M. Tan, 2014). At the same time, swimming of fish includes rhythmic activity, which is mainly produced by central pattern generators (CPG) at the level of the spinal cord (Ekeberg, 1993; Grillner et al., 1981)
The increased interest of the scientific community in the CPG is due to the fact that animals have excellent adaptability, flexibility and variability in their movements, which makes them the standard for performing robotic movements. (Yu. Ding, 2013). An increased efficiency of godlike underwater vehicles is also predicted in comparison with technical devices using propellers and standard control schemes for movement (Zhao, W., Hu, Y 2012, Fish, 2013). Due to the inherent non-linear dynamic properties of CPG models, abrupt changes in the angles of oscillatory joints can be avoided and a smooth transition between gaits can be achieved. Notably, CPGs eliminate the need for trajectory planning and precise knowledge of the characteristics of a mechanical system (Yu, J., Wen, L., 2017).

The most important discoveries and facts about fish swimming are that when a fish swims, there is a body wave moving along the body of the fish, and the direction of the body wave is opposite to the direction of its movement (K. Rossi, V. Coral, 2011; V. Lee, T . Wang, 2012).
In recent years, many robotic fish have appeared that imitate various types of fish and are based on this discovery (S. Zhang, Yu Qian, 2016). Based on this discovery, the work uses two main approaches to control the movement of robotic fish.
The first approach is based on the use of simple sinusoidal functions (J. Yu, L. Liu, 2006, 2008). Each joint of the robotic fish is assigned its own sinusoidal function. This approach is quite simple and easy to implement. However, it is not capable of handling complex environments or unpredictable effects. In addition, the sine wave does not correspond to the real form of movement of floating species (X. Niu, J. Xu, Q., 2014).

The biological mechanisms of fish swimming are used by artificial underwater systems based on different types of fish swimming. Tunimorphic swimming is typical of some of the fastest marine animals, such as the scombreeding, laminid sharks and cetaceans (Summers, 2004). These animals are physically characterized by a well streamlined body and high elongation, attached to the body by a narrow caudal peduncle, attached to the body by a narrow caudal peduncle. During swimming it is kept almost motionless, and significant lateral movements occur only in the region of the stem and tail. The thrust is created exclusively by the tail fin, which occurs like a flapping hydrofoil. Anguiliform swimming in this case shows a propagating wave that starts from the head of the fish and continues to the tail. As a result, the so-called "yaw" type of movement is formed.

1) movement of the body and / or caudal fin (BCF) (Q. Yan, Z. Han, S. Zhang and J. Yang, 2008; J. Liu and H.. Hu, 2010) and 2) the use of their median and pectoral fins (MPF) (C. Zhou and KH Low, 2012; T. Hu, KH Low, L. Shen and X. Xu, 2014)

Further we are presented on demonstrating undulating movements, i.e. lampreys, fish, amphibians, and snakes. Each separate section of a certain type. At the same time, the works are placed in the chronological order of their publication. At the end of each section, there is a summary that categorizes the articles described by the type of CPG used.

## Description of work related to the modeling of the CPG of lampreys

Elisa Donati, Federico Corradi, 2014 implemented a CPG network using neuromorphic electronic circuits that can be directly linked to the robotic actuators of a biomimetic robotic lamprey, eliminating the difference between software and hardware. These circuits are composed of analog silicon neurons and low power synapses.

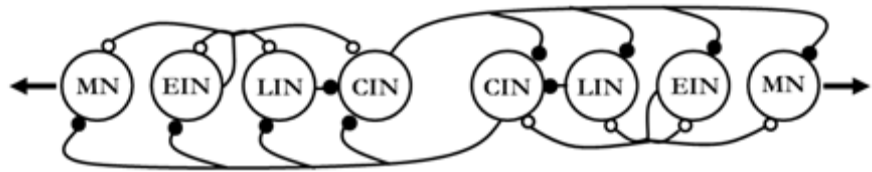

Fig. 2: Spinal cord CPG unit, comprising two sides with multiple types of neurons. The neurons are coupled by both excitatory and inhibitory connections. Filled circles represent inhibitory synapses while open circles the excitatory.

Pic. 1. Proposed architecture for CPG.

An on-chip oscillating network is used to provide the input signal to the model. This network consists of two pools (each with 4 silicon neurons). The connectivity within each pool is 0.5. The two pools suppress each other through inhibitory bonds with a connectivity coefficient $c = 0.5$. Then the authors built a hardware model of two segments of the CPG network, consisting of groups of neurons of types MN, EIN, LIN, CIN, as shown in Fig. 2. Each group of neurons consists of five connected neurons with a connectivity coefficient $c = 0.2$. A schematic diagram depicting a complete hardware model with both an input oscillatory network and CPG segments is shown in Figure 5.

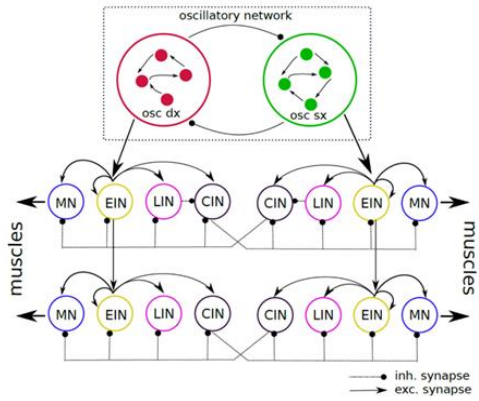

Fig. 5: Schematic diagram of the hardware model. The top box contains the oscillator network composed of two pools of neurons *osc dx* and *osc sx*. The bottom lines represent two CPG segments that couple the different types of neurons described in Section II-A

Excitatory interneurons (EIN) are projected onto all other neurons from the same side: lateral and contralateral inhibitory neurons (LIN and CIN, respectively) and motor neurons (MN). Motor neurons MN project the output signal directly onto the muscles. Contralateral inhibitory neurons are projected onto the contralateral side, and lateral inhibitory neurons are projected onto neurons on the ipsilateral side. Excitatory interneurons are thought to be responsible for the burst of activity.

The chip parameters were tuned to match those used in a biophysically realistic CPG control model to obtain rhythmic burst dynamics similar to those observed in a real lamprey CPG system. Experimental results have shown that the neuromorphic chip can mimic the behavior of the theoretical CPG model, offering the ability to directly control the actuators of an artificial bio-inspired lamprey robot.

**In 2018 C. L. Hamlet, et. all** investigated various functional forms of sensory feedback based on the curvature of the body and assessed their effect on sustainable energy and swimming kinematics, since little is known experimentally about the functional form of curvature feedback. The distributed CPG is modeled as a chain of connected oscillators. Pairs of phase oscillators represent the left and right sides of the segments along the lamprey body. They activate muscles that flex the body and move the lamprey through the fluid, which is modeled using the full Navier-Stokes model. The resulting curvature of the body then serves as an input to the CPG generators, closing the loop.
Model:

$$\frac{d\theta_{k,i}}{dt} = \omega + \sum_{j=1}^{n} \alpha_{i,j} \sin(\theta_{k,j} - \theta_{k,i} - \psi_{ij}) + \alpha_c \sin(\theta_{k,i} - \theta_{k^*,i} + \pi) + \eta(\kappa_i). \quad (1)$$

Here, $\theta_{k,i}$ represents the phase of the $i^{th}$ oscillator in the chain, on the $k^{th}$ side, where $k = 1$ represents the right side, and $k = 2$ represents the left side. The notation $k^*$ indicates the opposite side, in other words, if $k = 1$, then $k^* = 2$. The natural frequency of these oscillators is denoted by $\omega$ in Eq (1), and throughout this work we choose $\omega = 2\pi$ rad/sec.

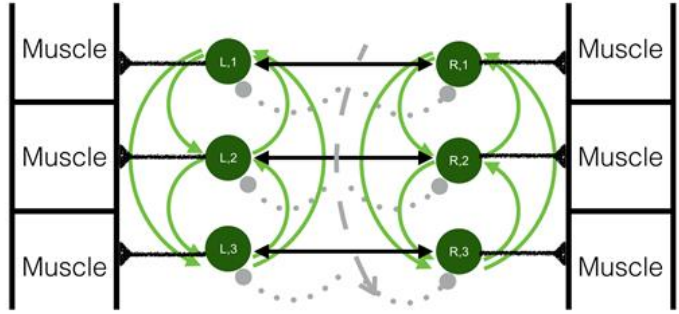

**Fig 3. Coupling of the oscillators.** To construct the phase oscillator model of the CPG, each oscillator is connected to every other oscillator on the same lateral ("L = left" or "R = right") side (solid green lines). Within one segment, numbered from head to tail, the left and right oscillators are also coupled together (solid black lines). The dotted gray lines represent curvature and its influence on the CPG oscillators. The curvature at each segment is calculated such that positive curvature curves in the direction of the right side measured from head to tail.

Muscles are activated by an electrical signal from the CPG. Our previous lamprey swimming model prescribed a wave of neural activation to mimic CPG input, but this neural activation was independent of evolving dynamics.

On the contrary, here we simulate the full Navier-Stokes hydrodynamics, in which the effects of the fluid on the developing curvature of the body, as well as the dynamics of the vortex going beyond the limits of the tail motion, are recorded. Although we are using a 2D fluid model, comparisons with robotic angel swimmers and real biological swimmers show excellent agreement on many flow parameters.

The CPG is modeled using a double circuit of sinusoidally coupled phase generators (one for the left side and one for the right, see Fig. 3):

In this work, the muscles of the lamprey are also modeled by using a kinetic model of the dynamics of calcium in each muscle and associating it with the Hill-type muscle force generation model, modified based on the data of experiments with lampreys.

Sensory feedback.

The CPG responds to sensory feedback. Here we simulate proprioceptive (perceiving body) feedback from edge cells. As discussed above, edge cells are mechanoreceptors that sense stretch along the body and send signals that serve to suppress the contralateral side of the body and excite the ipsilateral side (relative to the position of the edge cell) with increased stretch. Recent findings by Massarelli et al. showed that marginal cells also respond to the rate of change in extension. Here, for the sake of simplicity, we restrict our feedback model to only stretching, which is done by controlling the curvature of the body. Details regarding the functional form of input from border cells in the CPG are currently unknown. To represent the feedback from the edge cells to the CPG, we add a feedback term in equation (1) as $\eta(\kappa_i)$, where $\kappa_i$ is the curvature of the lamprey midline in segment i. As a starting point, we model this as an additive response. At each time step, the curvature is calculated along the midline of the body from head to tail at point $(x(s), y(s))$ using the formula:

$$\kappa = \frac{x'y'' - y'x''}{((x')^2 + (y')^2)^{3/2}} \quad (12)$$

The paper considered two different functional feedback forms:

$$\eta(\kappa_i) = \eta_m |\kappa_i|, \quad (13)$$

which we call magnitude feedback (denoted M), and

$$\eta(\kappa_i) = (-1)^k \eta_d \kappa_i, \quad (14)$$

which we call directional feedback (denoted D), and where $i$ is the segment number, $k$ is the side of the body (1 is left and 2 is right), and $\eta_m$ and $\eta_d$ are the constant gain parameters with units of cm rad/sec. We will discuss the implications of these feedback forms in the Results section below.

In addition to the CPG, the body of the lamprey with neutral buoyancy was modeled in the work, it supports passive forces due to elastic connections, as well as active forces on the lateral sides due to muscle contractions. The strength and timing of the contractions of individual muscle segments develop with CPG, calcium dynamics, and a Hill-type muscle model, each of which is associated with the evolving body shape. We adopt a submerged boundary structure that links the forces supported along the three filaments of the lamprey body, collectively denoted X (s, t), with the surrounding viscous incompressible fluid:

Due to the connection between the body and the fluid during modeling, a phase lag naturally occurs, in which muscle activation occurs after curvature of the body, even without feedback. This effect is due to the periodic nature of the floating oscillations and stabilizes several swimming cycles after starting the simulation. For the current model, the phase lag between muscle activation and curvature was calculated.

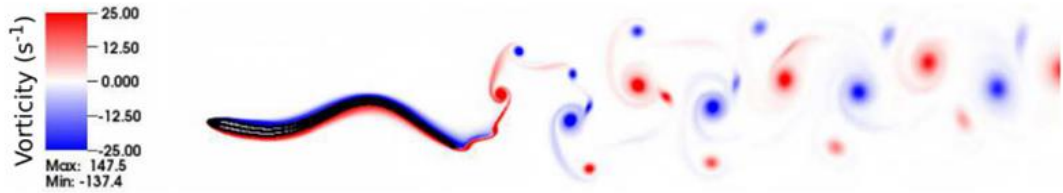

**Fig 7. Simulation with no feedback.** The body kinematics and wake structure for the control case is shown at $t = 8.0$ s. The outline and midline of the computational lamprey are shown in black. Vortices shed from the tail are shown in red (counterclockwise rotation) and blue (clockwise rotation). This model lamprey swims at a speed of 0.52 L · s$^{-1}$ with a tail beat amplitude of 0.12 L and a body wavelength of 0.75 L. Natural lampreys swim at about 0.1 L/s (body lengths per second) during migration [48] with maximum sustained speeds of about 2.5 L/s [49].

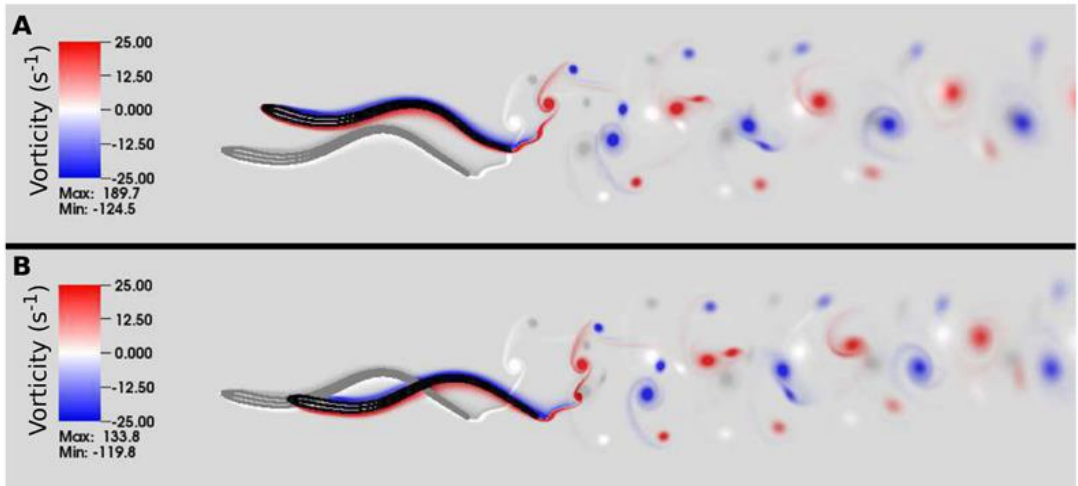

**Fig 9. Effects of directional feedback.** Body configuration and trailing wake structures for lamprey simulations after 8 s of simulated time. The control case (without feedback) is shown in gray. Panels **A** and **B** show the effects of feedback negative and positive gains ($\eta_d = -15$ cm · s$^{-1}$ and 15 cm · s$^{-1}$, respectively), and the corresponding dynamics are shown in S1 and S2 Movies.

The influence of feedback on the states of the model and the characteristics of its dynamics were considered.
The feedback model in this paper approximates two different known effects in the lamprey sensor motor.
control system: firstly, the phase effect of edge cells, which tends to increase the current locomotor rhythm, and, secondly, the general exciting effect of sensory feedback. We found that even in the absence of disturbances, both effects are important in steady motion. In our model, both forms of feedback primarily affect the duration and frequency of muscle contractions.

**In the article Emmanouil Angelidis, Emanuel Buchholz, 2021** a neuromechanical simulation was implemented, for which an accurate 3D model of a lamprey robot was used, which consists of nine body parts. These parts are connected by eight joints, which have one degree of freedom: rotation around the vertical axis. To create swimming models, the angular positions of these joints oscillate with amplitudes, frequencies and phases prescribed by the CPG model.

In this work, a double chain model is used, in which the activity of one side of the spinal cord is in antiphase with the activity of the other side, which is observed when measuring the muscle activity of lampreys. The control centers between the oscillators on each side can shift the overall oscillatory patterns that cause rotation due to changes in the overall curvature of the robot. This dynamic system is described by the following differential equations in phase space:

$$\dot{\theta}_i = 2\pi \nu_i + \sum_j r_j w_{i,j} \sin\left(\theta_i - \theta_j - \Phi_{i,j}\right) \tag{1}$$

$$\ddot{r}_i = a_i \left(\frac{a_i}{4}(R_i - r_i) - \dot{r}_i\right) \tag{2}$$

$$x_i = r_i \left(1 + \cos \theta_i\right) \tag{3}$$

$$\Psi_i = \alpha \left(x_{i,\text{right}} - x_{i,\text{left}}\right). \tag{4}$$

Θi, vi is the phase and preferred frequency of the ith oscillator, ri is the amplitude, xi is the output of the ith oscillator, which represents the activity of the motor neuron, and Ψi is the output of the model applied to the robot. and combines the activity of the oscillators on the left and right side of the double chain model.
The same model in Cartesian space:

$$\dot{x}_i = a(R_i^2 - r_i^2)x_i - \overline{\omega}_i y_i \tag{5}$$

$$\dot{y}_i = a(R_i^2 - r_i^2)y_i + \overline{\omega}_i x_i \tag{6}$$

$$\overline{\omega}_i = \omega_i + \sum_j \frac{w_{ij}}{r_i}[(x_i y_j - x_j y_i)\cos \Phi_{i,j} - (x_i x_j - y_i y_j)\sin \Phi_{i,j}] \tag{7}$$

where xi, yi denote the x and y coordinates of a point in 2D space moving in a circle in time, with a frequency controlled by equation (7). The parameter a determines the rate of convergence of the amplitude to the steady state, and ri is the norm of the vector [x, y]. This formulation is close to the standard form of coupled Hopf oscillators coupled with other oscillators. This equation has the advantage that the x, y values remain within the limit cycle, the radius of which is determined by the amplitude of the oscillations, solving the problem of continuously increasing phase when it is necessary to use the phase representation.

To enable drive corresponding to high-level stimulation, two piecewise linear functions are used, which saturate when the stimulation falls outside a certain range. These two functions control the target frequency and target amplitude of each oscillator according to the ratios:

$$\omega_i(d) = \begin{cases} c_{\omega,1}d + c_{\omega,0}, & \text{if } d_{\text{low}} \leqslant d \leqslant d_{\text{high}} \\ 0, & \text{otherwise} \end{cases} \qquad (8)$$

$$R_i(d) = \begin{cases} c_{R,1}d + c_{R,0}, & \text{if } d_{\text{low}} \leqslant d \leqslant d_{\text{high}} \\ 0, & \text{otherwise.} \end{cases} \qquad (9)$$

These two equations repeat biological observations that the frequency and amplitude of muscle contractions increase with increased stimulation. They complement the CPG with high-level modulation and provide a complete mathematical formulation of the control structure that is implemented in the SNN.

**Spike neural network model:**

SNN is a modular structure in which one vibrational center is represented by one population of impulse neurons and calculates equations (5) - (7). At the same time, this population encodes equation (9). As a connection between neural oscillators, an intermediate population is introduced, which receives x, y values from neighboring oscillators, and calculates the coupling term of equation (7). This intermediate population exchanges data between neural oscillators. In this case, each of the vibrational centers receives input data from a higher-level module through equations (8) and (9).

The building block of the CPG model is shown in Figure 3, which shows the neural engineering framework (NEF) model of a single Hopf-type generator. It is a neural network with one hidden feed-forward layer with three inputs (x, y and ω) and two outputs (τ (a (R2 - r2) x - ωy) + x and τ (a (R2 - r2) y + ωx) + y). The weights for this network were found by randomly sampling the inputs (x, y, and ω), calculating the desired output for each input, and then optimizing the network with that data in mind.

The modeling in the article was carried out using the Nengo platform, which has built-in methods for generating neural networks that approximate differential equations, and the simplest computational model of impulse neurons LIF, which are combined in populations for each phase oscillator.

Results:

This article presents a central pattern generator (SCPG) based on a high-level system of abstract coupled Hopf-type oscillators. The method allows smooth control of a lamprey robot that adapts to various simulation scenarios with high-level drive control. The authors brought simulations under the NRP to perform several locomotive tasks. In comparison with other models, this model, on the one hand, requires more neurons, but on the other hand provides richer dynamics. A limitation of this approach is that the spikes generated by the neural network are not directly used to control the robot. Another limitation is that to accurately reproduce a mathematically formulated system, a large number of neurons are required to reduce noise.

Stimulation of one of the oscillators temporarily causes its disturbance, as well as adjacent oscillators, which quickly recover from the disturbance. This was demonstrated in the work under the following scenarios:

(a) Perturbation of a single vibrational center by external stimulation by adding an additional time term to one of the oscillators in equation (5)

(b) Asymmetric spinal cord stimulation from left to right of the spinal cord

These scenarios show the ability of the SNN model to quickly recover from external disturbances, as well as modulate swimming gait.

The authors applied the developed system to simulate interactions using the neuro-robotics platform (NRP).

## Description of works related to modeling the CPG of fish

The CPG of swimming movements can be subdivided into CPG with an open and closed loop, as described by Korkmaz 2018. In this case, an open loop is used to form various swimming patterns. Closed-loop CPG uses feedback to adapt behavior to a real environment (imitation of the real mechanism of animal CPG activity.

Attempts to reproduce the rhythmic movements underlying the movement of fish over long distances were the first stage in modeling the movement of fish. In Zhang et. all. 2006 proposed a scheme in which the control loop is a series-connected blocks of two mutually inhibiting each other by means of the CPG feedback. The main role in determining the type of movement and its adaptation to changing external conditions is played by the incipient inputs from the brain tissues. It is important to note that downward connections from the brain go to each segment of the CPG. When one of the CPGs in the segment is activated, the muscle for which he is responsible is shredding, while the muscle of the paired CPG relaxes. Stretching the muscle causes the activation of the receptor, the action of which inhibits the activity in the previously triggered CPG and stimulates the activation of the paired generator. Thus, rhythmic sequential movements are obtained. Intersegmental delays are kept constant to maintain a stable waveform, but the intra-segment delay between activation of paired CPGs can be different to simulate different movement speeds.

In this work, the authors refused to use the Matsuoko oscillators, because in this model, it is impossible to modulate the parameters of oscillators autonomously, due to nonlinear interactions between them. Oscillator Zhang is proposed, devoid of the indicated drawback:

$$\begin{cases} \dot{u} = \omega v + f(u) \\ \dot{v} = -\omega u + f(v) \end{cases},$$

where u and v 'denote the inhibitory and excitatory activities of neurons, respectively, w is a parameter that mainly determines the angular frequency of oscillations, and f (x) is a nonlinear function. If we set the non-linear function to zero for all time, then the system becomes a total sine-cosine oscillation, which is written as:

$$\begin{cases} \dot{u} = \omega v \\ \dot{v} = -\omega u \end{cases}.$$

Also, the Zhang oscillator has many convenient control properties for decomposing parameters into a nonlinear function, in contrast to the composition of the Wilson-Cowen oscillator. This makes it possible to modulate each vibration parameter. The fish is capable of starting, stopping, swimming forward, backward and turning **(Zhang, 2006).**

**Another frequently cited work** is an article by Lachat, 2006 Pltcm robot control architecture is built around a central pattern generator (CPG) implemented as a system of coupled nonlinear oscillators that, like its biological counterpart, can create coordinated patterns of rhythmic activity modulated by simple control parameters.

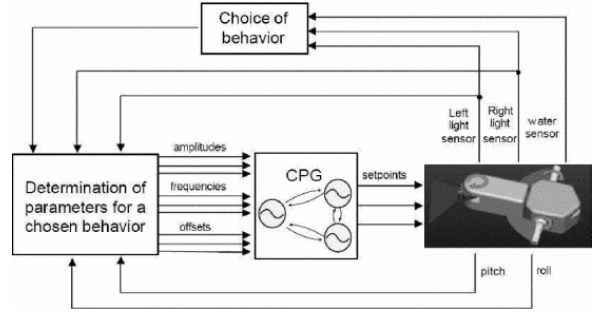

Using the CPG model, the robot can perform and switch between different types of locomotor behaviors such as swimming forward, swimming backward, turning, rolling, moving up / down, and crawling. This behavior is triggered and modulated by sensory input from light and water sensors.

The CPG is implemented as a system of three coupled amplitude-controlled phase oscillators, one per edge. Oscillator i is implemented as follows:

$$\dot{\phi}_i = \omega_i \qquad (1)$$
$$+ \sum_j (\alpha_{ij} r_j \sin(\phi_j - \phi_i) + \beta_{ij} r_j \cos(\phi_j - \phi_i))$$

$$\ddot{r}_i = a_r(b_r(R_i - r_i) - \dot{r}_i) \qquad (2)$$
$$\ddot{x}_i = a_x(b_x(X_i - x_i) - \dot{x}_i) \qquad (3)$$
$$\theta_i = x_i + r_i \cos(\phi_i) \qquad (4)$$

This CPG model has several good properties. The first interesting property is that the system exhibits limit cycle behavior. The second interesting property is that the control parameters ωi, Ri and Xi can change abruptly and / or constantly, causing only smooth modulation of the setpoint fluctuations (i.e., no jumps or jumps).

Another interesting feature is that feedback terms can be added to equations 1–3 to maintain synchronization between control oscillations and mechanical oscillations, which will be considered in future works.

**In 2008 Hu and Zhao** proposed their own type of CPG, also based on Matsuoko's neurons. Unlike previous work, in this study, the generator segment is a neural oscillator of two neurons connected by mutual inhibitory connections and responsible for the flexor and extensor. The model equations are optimized so that the oscillator operates in the limit cycle mode:

$$T_u \dot{u}_i^e = -u_i^e - \beta v_i^e - \alpha y_i^f - \sum_{j=1}^n \omega_{ij} y_j^e - Feed_i^e + u_i^{0e} \qquad (1)$$

$$T_v \dot{v}_i^e = -v_i^e + y_i^e \qquad (2)$$

$$T_u \dot{u}_i^f = -u_i^f - \beta v_i^f - \alpha y_i^e - \sum_{j=1}^n \omega_{ij} y_j^f - Feed_i^f + u_i^{0f} \qquad (3)$$

$$T_v \dot{v}_i^f = -v_i^f + y_i^f \qquad (4)$$

$$y_i^{\{e,f\}} = max(u_i^{\{e,f\}}, 0) \qquad (5)$$

$$y_i = y_i^f - y_i^e \qquad (6)$$

To simulate the movement of a fish, each of the joints is assigned its own oscillator. In this case, its output signal is converted into an angular characteristic for the motor responsible for this segment. In this model, the movement involves not only the joints of the body, but also the fins. The authors proposed an approach whereby different types of roll can be modeled by presetting different weights between the neural oscillators. The model demonstrates the ability to swim with or without fins. The fish is capable of moving forward, backward, making a sharp turn, diving and lifting, and braking.

**In 2012** Yu, Ding et all presented an amphibious robot capable of movement on land and swimming. For swimming controls, a CPG based on oscillators (J. A. Acebron, L. L. Bonilla, 2005) was used, whose amplitude varied within the Kuramoto model (R. Ding, J. Yu, 2010).

$$\begin{cases} \dot{\phi}_i = 2\pi f_i + \sum_{j \in T(i)} a_j w_{ij} \sin(\phi_j - \phi_i - \gamma_{ij}) \\ \ddot{a}_i = \tau_i \left\{ \frac{\tau_i}{4}(A_i - a_i) - \dot{a}_i \right\} \\ \chi_i = a_i \left\{ 1 + \cos(\phi_i) \right\}. \end{cases}$$

The CPG consisted of 6 segments, each of which included two mutually suppressed oscillators responsible for the flexor and extensor muscles. In this case, the change in the position of the segment was determined by subtracting the signals received from the oscillators. The authors point out that, in contrast to the scheme proposed by Ijspreet (Ijspeert, A, 2007), in the new model there is a piecewise-linear function that determines the contribution of a separate CPG segment depending on the speed of fish movement:

$$\begin{cases} \dot{\phi}_i = 2\pi f_i + \sum_{j \in T(i)} a_j w_{ij} \sin(\phi_j - \phi_i - \gamma_{ij}) \\ \ddot{a}_i = \tau_i \left\{ \frac{\tau_i}{4}(A_i - a_i) - \dot{a}_i \right\} \\ \chi_i = a_i \left\{ 1 + \cos(\phi_i) \right\}. \end{cases}$$

where d indicates the input control signal received by the robot, which can be further split into left and right control signals dL and dR associated with bidirectional neural oscillators. flow-cut and Alow-cut represent the frequency and amplitude of the low cut when oscillating is blocked, respectively. kf, i, bf, i, kA, i, bA, i denote frequency coefficients and amplitude coefficients that determine the evolution of the natural frequency and amplitude of the i-th oscillator. dlow, i and dhigh are the lower and upper thresholds separately. For the i-th oscillator, as soon as the drive reaches the corresponding lower threshold dlow, i, homologous oscillations begin. By adjusting the lower threshold dlow, i, the corresponding hinge can be enabled or disabled (for example, dlow, i> dhigh), which can realize various driving modes coordinated by a plurality of driving elements. The same is the case with the pectoral fins, which are denoted dlow, pec. The authors argue that such flexible control policies endows the robot with energy-efficient swimming.

**Continuation of this work (Yu, Ding) was a study** published in 2013, where the authors modernize the model in order to provide the possibility of transitions between patterns. In addition, the model has integrated CPG-based sensory feedback to modulate the phase displacements of contralateral oscillators for different patterns.

$$\begin{aligned} \dot{\theta}_i &= 2\pi f_i + \sum_{j \in T(i)} a_j w_{ij} \sin(\theta_j - \theta_i - \phi_{ij} + u_i), \\ \ddot{a}_i &= \tau_i \{ \tfrac{\tau_i}{4}(A_i - a_i) - \dot{a}_i \}, \\ x_i &= e_i + a_i \cos \theta_i, \end{aligned}$$

where  and  are the state variables of the ith oscillator, representing the phase and the amplitude, respectively.  and  denote the intrinsic frequency and amplitude of the oscillator, respectively.  is a positive time constant determining how quickly the amplitude variable  converges to .  and  indicate the coupling weight and phase bias of the jth oscillator to the ith oscillator.  is the discrete set of oscillators exerting inbound couplings on the ith oscillator.  represents the extracted output signal adjusted by an oscillation offset . In particular, like , satisfies . stands for the feedback signal related to the ith oscillator. It should be remarked that the feedback signal is usually an integral component of the differential equations that govern the CPG (Righetti and Ijspeert, 2008; Seo et al., 2010).

The CPG-based control in our previous work is an open loop, and motion control in surface and underwater conditions is carried out separately (Yu et al., 2012); In this article, we use sensory feedback from liquid level sensors to create a closed loop. Specifically, two liquid level sensors (see Figure 2) are used to detect the presence / absence of water to determine if the robot is on land or in water. The presence / absence of water information is then included in the CPG to explicitly modulate the oscillator coupling phase shifts to generate reactive behavior.

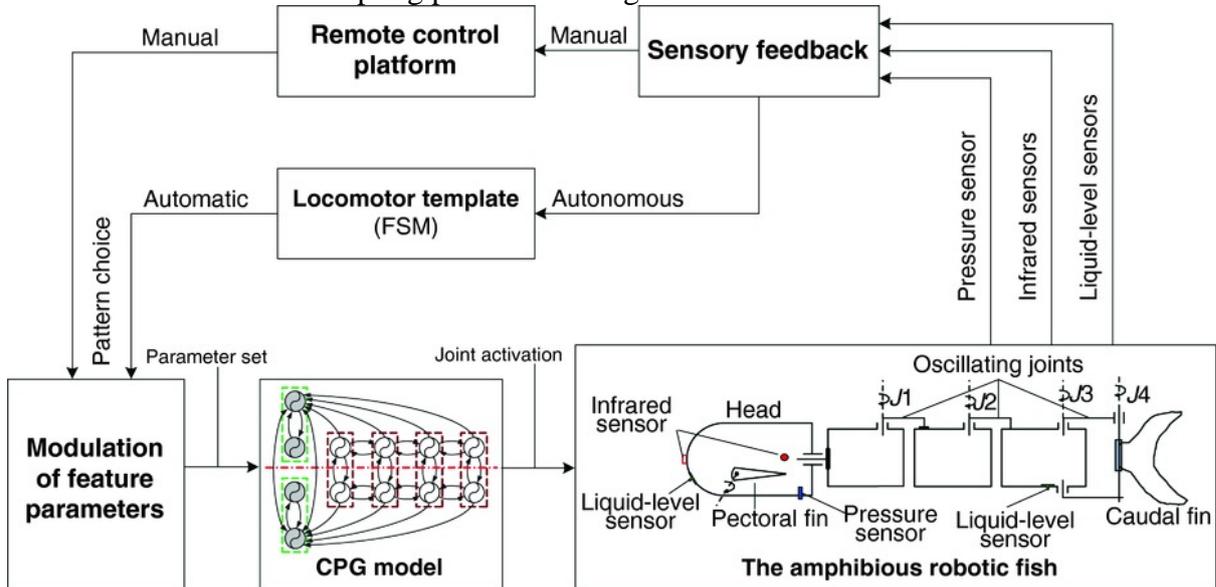

The fish is capable of moving forward. back, turn and turn in place. Of particular note is the possibility of modeling vertical swimming like dolphins.

**In 2013, Wang and Xie (CLOSED CPG)** introduced a model based on Crespi oscillators:

$$\ddot{a}_i = \alpha\left(\alpha(A_i - a_i) - 2\dot{a}_i\right) \quad (4a)$$

$$\ddot{x}_i = \beta\left(\beta(X_i - x_i) - 2\dot{x}_i\right) \quad (4b)$$

$$\ddot{\phi}_i = \sum_{j=1, j\neq i}^{N} \mu_{ij}\left(\mu_{ij}a_j(\phi_j - \phi_i) - 2(\dot{\phi}_i - 2\pi f_i)\right) \quad (4c)$$

$$\theta_i = x_i + a_i \cos(\phi_i) \quad (4d)$$

where ai, xi and φi are state variables representing the amplitude, displacement and phase of the ith oscillator, and variable θi is its output. The fi, Ai and Xi parameters are control parameters for the desired frequency, amplitude and vibration offset. $\mu_{ij}$ and $\phi_{ij}$ determine the coupling weight and phase displacement of the j-th generator to the i-th generator. $\alpha_i$ and $\beta_i$ are structure parameters representing the dynamic characteristics of the ith oscillator. $T_i$ is a set of neighbors of the i-th oscillator that interact with the i-th oscillator.

In this case, the parameters of the model are modified so that the oscillators are in the state of the limit cycle.

The presented CPG model consists of three functional layers, namely the input saturation functions, coupled neural oscillators and the output transient function.

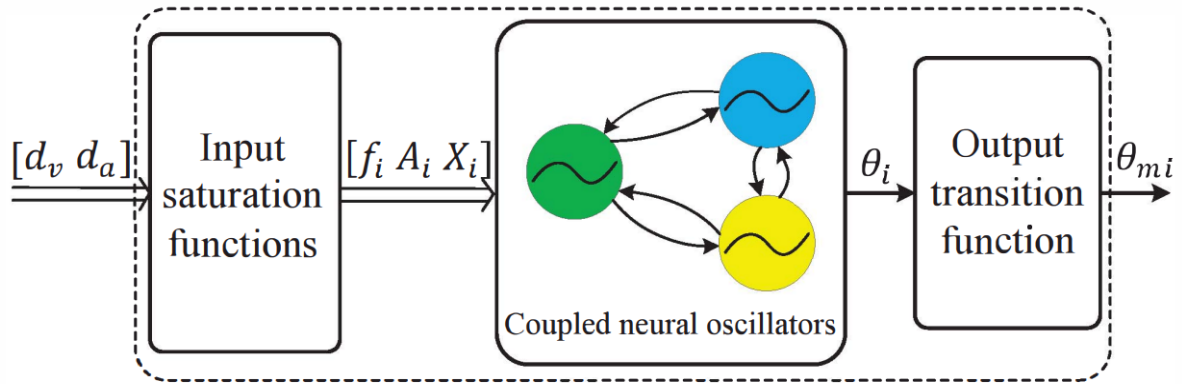

Fig. 3. The proposed three-layer CPG model.

The saturation function receives command signals to start, regulate, and stop CPG from centers like the brainstem. The $d_v$ command defines the speed and transition of the robot to gait using two input saturation functions, the $d_a$ command defines several swimming modes such as forward swim, pivot and roll. There are N associated oscillators, functionally similar to the spinal cord in the biological nervous system, which generates rhythmic movement patterns. The parameters $f_i$, $A_i$ and $X_i$ are the frequency, amplitude and displacement of the i-th oscillator, respectively. Similar to muscles that respond to neural signals, the output transition function converts the output signal $\theta_i$ of the oscillatory network into a motor drive signal $\theta_{mi}$.

where $d_v$, min and $d_t$, max represent the lower and upper $d_v$ thresholds of the i-th generator, respectively; coefficients ($c_{Ii}$, $CA_i$), offset ($f_{i.h}$, $A_{i.h}$) and saturation ($f_{i.sat}$, $A_{i.sat}$) are used to modulate the frequency and amplitude of the i-th oscillator, respectively.

$$f_i = g_{fi}(d_v) = \begin{cases} c_{fi}d_v + f_{i,b}, & \text{if } d_v^{i,\min} \leq d_v \leq d_v^{i,\max}; \\ f_{i,sat}, & \text{otherwise.} \end{cases} \quad (1)$$

$$A_i = g_{Ai}(d_v) = \begin{cases} c_{Ai}d_v + A_{i,b}, & \text{if } d_v^{i,\min} \leq d_v \leq d_v^{i,\max}; \\ A_{i,sat}, & \text{otherwise.} \end{cases} \quad (2)$$

$$X_1 = g_{X1}(d_a) = \begin{cases} \pi d_a & \text{if } d_a \in \{-\frac{1}{2}, \frac{1}{2}\} \\ 0, & \text{otherwise.} \end{cases} \quad (3a)$$

$$X_2 = g_{X2}(d_a) = \begin{cases} \pi(d_a - sgn(d_a)) & \text{if } d_a \in \{-\frac{1}{2}, \frac{1}{2}\} \\ 0, & \text{otherwise.} \end{cases} \quad (3b)$$

$$X_3 = g_{X3}(d_a) = \begin{cases} \pi d_a & \text{if } -\frac{1}{3} \leq d_a \leq \frac{1}{3} \\ 0, & \text{otherwise.} \end{cases} \quad (3c)$$

The third level is the output transient function hi (8i), which converts the signal of the rhythmic neural oscillator into a signal to turn on the motor. It takes the form:

$$\theta_m^i = h_i(\theta_i) = \begin{cases} \theta_m^{i,\max}, & \theta_i > \theta_i^{\max} \\ c_{mi}\theta_i + \theta_m^{bi}, & \theta_i^{\min} \leq \theta_i \leq \theta_i^{\max} \\ \theta_m^{i,\min}, & \theta_i < \theta_i^{\min} \end{cases} \quad (5)$$

where θim is the control signal supplied to the corresponding servo motor; θbiim is the potential signal when the i-th servomotor remains in the home position and Cmi is the translation ratio.

**The continuation of the described work (HERE ALREADY CLOSED CPG) was published in 2014.** Wang and Xie also used the Crespi Oscillators, specifying that the indices i = 1, 2, 3 in this article denote the left pectoral fin, right pectoral fin and tail fin of the robot fish, respectively. The authors modeled the movement of ostracoid fish.

Now the CPG is two-tier. The first level of the CPG directly controls the movement of the fins. The oscillators of this layer are interconnected and each receives signals from the upper level of the CPG, which determine the types of swimming. This signal can be split into four command inputs, namely dv, dy, dp and dr. ... The dv command determines the speed of the robot, while dy, dp and dr adjust the yaw, roll and roll movements of the robot, respectively:

$$f_i = \begin{cases} c_{f,i}^{PF} d_v + f_{i,b}^{PF}, & \text{if } d_v^{i,\min} \leq d_v < d_v^{trans}; \\ c_{f,i}^{PBC} d_v + f_{i,b}^{PBC}, & \text{if } d_v^{trans} \leq d_v \leq d_v^{i,\max}; \\ f_{i,sat}, & \text{otherwise.} \end{cases} \quad (2)$$

$$A_i = \begin{cases} c_{A,i}^{PF} d_v + A_{i,b}^{PF}, & \text{if } d_v^{i,\min} \leq d_v < d_v^{trans}; \\ c_{A,i}^{PBC} d_v + A_{i,b}^{PBC}, & \text{if } d_v^{trans} \leq d_v \leq d_v^{i,\max}; \\ A_{i,sat}, & \text{otherwise.} \end{cases} \quad (3)$$

$$X_1 = \begin{cases} d_p + \text{sgn}\left(1 + (-1)^{\text{sgn}(d_p)}\right) d_r, \\ \text{if } d_p \in [-0.5\pi, 0] \cup [0, 0.5\pi] \cup \{\pi, -\pi\}, \\ d_r \in \{-0.5\pi, -\pi, 0.5\pi, \pi\}; \\ 0, \text{ otherwise.} \end{cases} \quad (4a)$$

$$X_2 = \begin{cases} d_p + \text{sgn}\left(1 + (-1)^{\text{sgn}(d_p)}\right)(d_r - \pi\text{sgn}(d_r)), \\ \text{if } d_p \in [-0.5\pi, 0] \cup [0, 0.5\pi] \cup \{\pi, -\pi\}, \\ d_r \in \{-0.5\pi, -\pi, 0.5\pi, \pi\}; \\ 0, \text{ otherwise.} \end{cases} \quad (4b)$$

$$X_3 = \begin{cases} d_y, & \text{if } d_y \in [-\pi/3, \pi/3]; \\ 0, & \text{otherwise.} \end{cases} \quad (4c)$$

where $\text{sgn}(\cdot)$ is a signum function and is defined as follows:

$$\text{sgn}(t) = \begin{cases} -1, & \text{if } t < 0; \\ 0, & \text{if } t = 0; \\ 1, & \text{otherwise.} \end{cases} \quad (5)$$

In this case, each of the oscillators receives input commands. This scheme allows not only to perform the basic types of movement of fish and adjust the speed of the robot, but also to switch swimming modes without significant delays.

It should be noted that this system, due to the presence of feedback, is capable of self-stabilizing the position of the robot without the intervention of the upper level of the CGR.

**Also in 2014, Niu, Xu, 2014** presented a completely new form of the CPG model, which consists of coupled Andronov-Hopf oscillators, an artificial neural network (ANN), and an external amplitude modulator. Unlike previous works, which only use coupled generators and therefore only fixed waveforms can be generated, the added ANN and an external amplitude modulator to the CPG structure allows different kinds of signals to be generated. Also, the design of coupled generators uses a three-dimensional topology, and higher compression rates can be achieved compared to those using traditional one-dimensional or two-dimensional topologies.

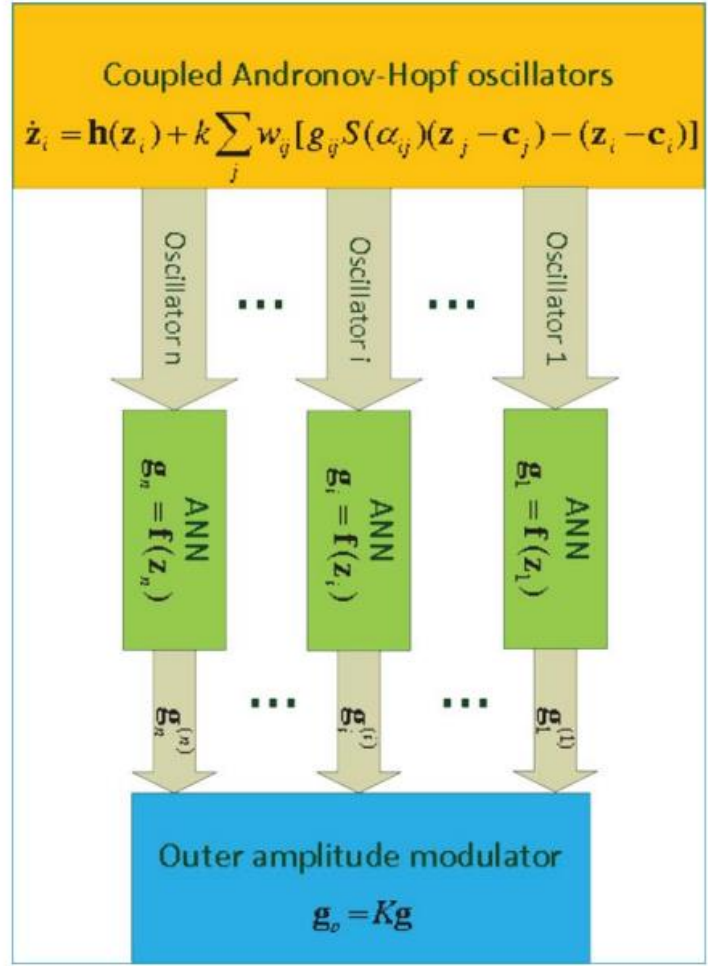

As shown in fig. 1, the CPG structure contains three main components. The first component is coupled Andronov-Hopf oscillators, which consist of several single Andronov-Hopf oscillators. In this case, the description of a single oscillator is not enough, and when they are combined into a single system, they take the following form:

$$\dot{z}_i = \mathbf{h}(z_i) + k \sum_j w_{ij} \left[ g_{ij} S(\alpha_{ij})(z_j - c_j) - (z_i - c_i) \right] \quad (2)$$

where i and j are ordinal numbers of oscillators, and oscillator j is an oscillator that is directly connected to oscillator i. The dynamics zi consists of two terms. The first term h (zi), describing the effect of zi itself, is the same as in the single Andronov - Hopf oscillator. The second term describes the relationship with other connected oscillators. k is the constant strength of the coupling, wij is the weight of the coupling between the two oscillators, gij = ai / aj is the ratio of the amplitudes of the two oscillators, αij is the desired phase difference between the two oscillators, ci is the center of oscillations, and S (αij) is the rotation transformation, i.e. e.

$$S(\alpha_{ij}) = \begin{bmatrix} \cos \alpha_{ij} & -\sin \alpha_{ij} \\ \sin \alpha_{ij} & \cos \alpha_{ij} \end{bmatrix}.$$

where z = [m, n] T is the state vector of a single Andronov – Hopf oscillator, c = [c1, c2] T is a constant vector representing the center of oscillations, a> 0 is the oscillator amplitude, β is the rate of attraction of oscillations, and ω> 0 - vibration frequency. The parameters c, a, β and ω can be adjusted according to the needs.

In fig. shows three types of CPG network topology. All three topologies obey the connection rule: each oscillator is connected only to the nearest generator (s).

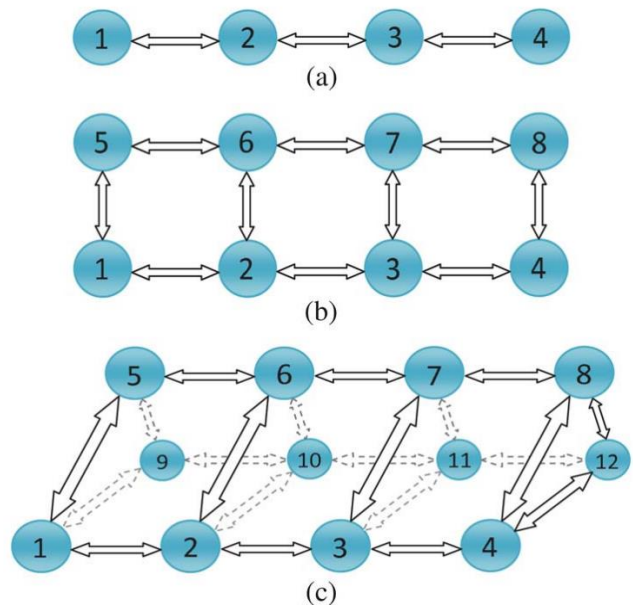

The authors demonstrate that in a three-dimensional system, reaching the limit cycle state is achieved faster due to the presence of a greater number of connections between the oscillators. In addition, the resulting pattern is more robust. The second component of the CPG is the ANN. After training on the target values and receiving the excitation signals from the coupled oscillators, the ANN will output our desired waveform patterns. ANN expression:

$$\mathbf{g}_i = \mathbf{f}(\mathbf{z}_i)$$

where i = 1, ..., n; zi is the state vector of the i-th Andronov-Hopf oscillator, which also serves as an input for the i-th ANN; g - ANN output; and f is a non-linear mapping. The periodic signal zi, which is obtained from coupled generators, can provide stable signals to drive the CPG.

Using the level responsible for modulating the signal, you can achieve the required signal amplitudes on each individual segment.

The authors study the principles of movement of real fish and, having found out the parameters, apply them to the created model. As a result, a control contour was obtained that can simulate the movement of real Anguilliform fish.

The article Hu, Liang 2015 built a CPG network based on Hopf oscillators with an explicit parameter controlling the phase difference between CPG blocks.

$$\begin{cases} \dot{u} = (\rho - r^2)u - \omega v, \\ \dot{v} = (\rho - r^2)v + \omega u, \end{cases}$$

A phase adjustment mechanism is proposed to compensate for phase errors in a physical system for controlling the movement of a robotic fish. Usually, the coupling term from one oscillator to another is realized as an additive perturbation that affects the phase dynamics of the forced oscillator, leading to frequency synchronization and stable phase shifts between them. However, the oscillation amplitude of the forced generator will also be changed, which makes it difficult to set the parameter for the desired oscillation amplitude. To eliminate this unwanted side effect, the disturbance signal p must act on the state variables of the forced generator as follows:

$$\begin{cases} \dot{u} = (\rho - r^2)u - \omega v + \dfrac{\epsilon p v^2}{r}, \\ \dot{v} = (\rho - r^2)v + \omega u - \dfrac{\epsilon p u v}{r}, \end{cases} \quad (6)$$

where ϵ > 0 means constant bond strength. Assuming u = rcos ϕ and v = rsin ϕ, equation (8) can be transformed into polar coordinates:

$$\begin{cases} \dot{r} = (\rho - r^2)r, \\ \dot{\phi} = \omega - \epsilon p \sin \phi. \end{cases} \quad (7)$$

The CGR network consists of two oscillators connected by downward links. It is assumed that instead of instantaneous communication, the generators interact through time delay communication. So the control oscillator can be modeled with:

$$\begin{cases} \dot{u}_2 = \left(\rho_2 - r_2^2\right)u_2 - \omega_2 v_2 + \dfrac{\epsilon u_1(t-\tau)v_2^2}{r_2}, \\ \dot{v}_2 = \left(\rho_2 - r_2^2\right)v_2 + \omega_2 u_2 - \dfrac{\epsilon u_1(t-\tau)u_2 v_2}{r_2}, \end{cases} \quad (8)$$

where $\tau$ is the communication delay, indices 1 and 2 are for the variables of the control and controlled oscillator, respectively. With this setting, the boost generator can be phase locked with the boost delay. The state variable delay corresponds to some phase shift in polar coordinates.

Adjusting the CPG parameters allows the system to quickly and smoothly switch to new limit laps. The desired phase difference was accurately demonstrated by the two gaits, thus confirming the effectiveness of the proposed connection scheme. The fish is capable of reaching speeds of 2 m per second (7.2 km / h). The fastest result that has been achieved.

**Although a CPG controller design using coupled oscillators** has been proposed previously, it cannot fully reproduce various rhythmic movements along with a smooth transition to simulate the versatility of animal movement. To address this problem, the authors propose a generic CPG model with an emphasis on its robustness analysis, smooth transition, and implementation architecture (**C. Wang, G. Xie, L. Wang and M. Gao, 2011**). The global exponential robustness of the model is derived using rigorous mathematical analysis. The transitions between different waveforms are also smooth, and the implementation architecture has low computational costs, therefore, it is suitable for microcontrollers. Moreover, all control parameters not only have an explicit link to the physical outputs, but can also be changed online. One of the key features of this CPG model is the use of partially linearized oscillators, which require low computational costs, while the performance of our model is satisfactory. Moreover, the structural parameters in our model have an explicit relationship with physical outputs and thus can be chosen more intelligently and easily in accordance with the requirements of the dynamic performance of the CPG.

In accordance with the three-link motor configuration, the architecture of the CPG network, consisting of three oscillators in this study (**J. Yu, C. Wang and G. Xie, 2016**), is illustrated in Fig.

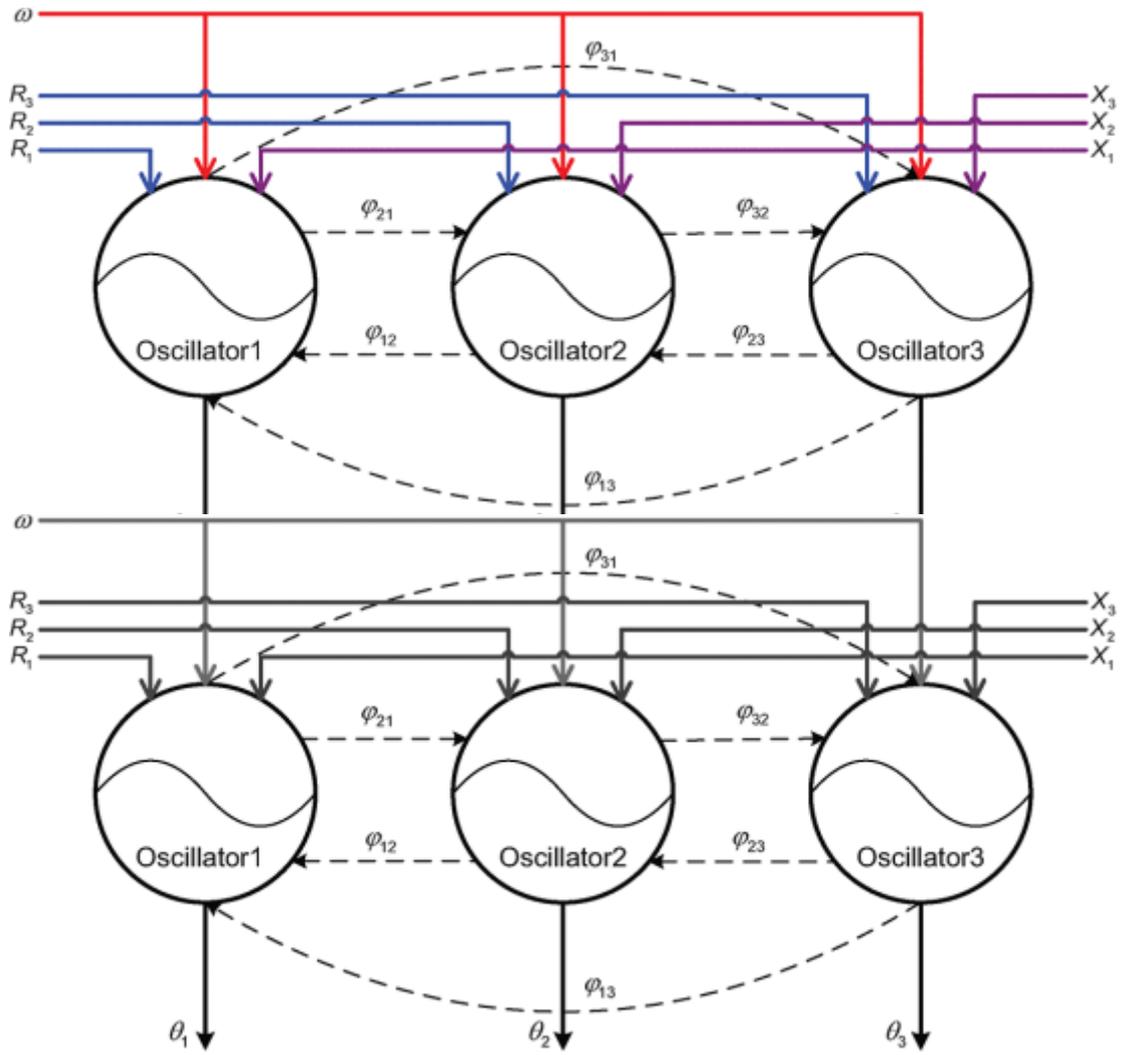

The black dotted line indicates the phase shift φij between two connected oscillators. The I-th oscillator has an input Ri in amplitude and an input Xi of an offset, described by blue and purple lines, respectively. All frequency inputs are the same, they are represented by red lines. In practice, the frequency is determined by the entire CPG network when it stabilizes. The black solid line next to θi indicates the output of the ith oscillator.

The mathematical model of the oscillator is expressed as follows:

$$\ddot{r}_i(t) = \alpha_i[\alpha_i(R_i - r_i(t)) - 2\dot{r}_i(t)]$$
$$\ddot{x}_i(t) = \beta_i[\beta_i(X_i - x_i(t)) - 2\dot{x}_i(t)]$$
$$\ddot{\phi}_i(t) = \sum_{j=1, j \neq i}^{N} \mu_{ij} \left[ \mu_{ij}(\phi_j - \phi_i - \varphi_{ij}) - 2(\dot{\phi}_i - 2\pi\omega) \right]$$
$$\theta_i(t) = x_i(t) + r_i(t)\cos(\phi_i(t))$$

In practice, the presented CPG model can be implemented as a four-part system, including an amplitude regulator in (1), a displacement regulator in (2), a phase and frequency regulator in (3), and an output quantity calculator in (4). The phase of each oscillator is influenced by other oscillators, while other variables, including the frequency described in (3), are independent. Finally, (4) combines all four parameters to generate sinusoidal signals that ultimately reproduce the backward propagating wave of the fish body.

This work is interesting in that it also provides a learning-based approach to the formation of interaction patterns between several model fish. However, since this deviates somewhat from the subject of the review, these methods will not be described here.

**In the next work, the authors also** sought to reduce computational costs. **S. Zhang, Y. Qian, 2016** pointed out that the developed CPG model must satisfy several requirements: (1) nonlinear equations must be simple to reduce computational costs for an inexpensive microprocessor; (2) the definition of the frequency, amplitude and phase difference of the output signals must be clear so that the transition of the control signals can be easily detected; (3) the CPG network must be able to receive feedback signals from IR sensors to achieve autonomous swimming.

In this study, the authors propose using two tails in a robot fish to use opposite forces to self-stabilize the mechanism. The robotic fish combines the characteristics and mechanisms of insect flight and fish swimming. A double-tail fin movement mechanism that transcends nature has been proposed to improve motor function and stability while swimming. Two pterygoid pectoral fins were equipped to allow vertical maneuvering. The matching CPG model has been designed to control swimming movements and smooth transitions.

This model uses oscillators and assumes that their output is a neuronal signal. The authors believe that ri → = (xi, yi) where ri → represents the state vectors of two neurons: xi and yi. For simplicity and low computational cost, a simple linear differential equation is used to describe neurons.

$$\begin{cases} \dot{x}_i = -\omega_i \cdot y_i \\ \dot{y}_i = \omega_i \cdot x_i \end{cases}$$

where ωi denotes the angular frequency of the ith system, and ωi = 2πfi> 0, ωi∈R, i = 1,2, .., n. The transformation of this system into polar coordinates and its solution leads the system to the following form:

$$\begin{cases} \dot{x}_i = \dot{r}_i \cos\theta_i - \dot{\theta}_i r_i \sin\theta_i = \frac{\dot{r}_i}{r_i} x_i - \dot{\theta}_i y_i \\ \dot{y}_i = \dot{r}_i \sin\theta_i + \dot{\theta}_i r_i \cos\theta_i = \frac{\dot{r}_i}{r_i} y_i + \dot{\theta}_i y_i \end{cases}$$

where $\dot{r}_i/r_i = 0.\ \theta$ indicates the phase of the oscillator. The solution shows that the outputs of the neurons are harmonic oscillations with a fixed amplitude.

The authors point out that when the amplitude changes, there are some constraints and the function h (ri; Ri; τi) that can satisfy the constraints is defined as shown below:

$$\frac{\dot{r}_i}{r_i} = h_{(r_i, R_i, \tau_i)} = \frac{\tau_i(R_i - r_i)}{r_i}$$

where τi> 0 (i = 1; 2; 3; ::: n) denotes a parameter that controls the convergence rate. Thus, the modified oscillator can be expressed by the following equations:

$$\begin{bmatrix} \dot{x}_i \\ \dot{y}_i \end{bmatrix} = \begin{bmatrix} \frac{\tau_i(R_i - r_i)}{r_i} & -\dot{\theta}_i \\ \dot{\theta}_i & \frac{\tau_i(R_i - r_i)}{r_i} \end{bmatrix} \cdot \begin{bmatrix} x_i \\ y_i \end{bmatrix}$$

coupled oscillators i and j can be expressed by the following equations:

$$\begin{bmatrix} \dot{r}_i \\ \dot{\theta}_i \end{bmatrix} = \begin{bmatrix} \frac{\tau_i(R_i - r_i)}{r_i} \\ \omega_i - r_i r_j \sin(\theta_i - \theta_j - \phi_{ij}) \end{bmatrix}$$

and

$$\begin{bmatrix} \dot{r}_j \\ \dot{\theta}_j \end{bmatrix} = \begin{bmatrix} \frac{\tau_j(R_j - r_j)}{r_j} \\ \omega_j - r_j r_i \sin(\theta_i - \theta_j - \phi_{ij}) \end{bmatrix} \quad (9)$$

The figure illustrates the relationship of Oscillator i with other oscillators. Equation 10 describes an oscillator in polar coordination.

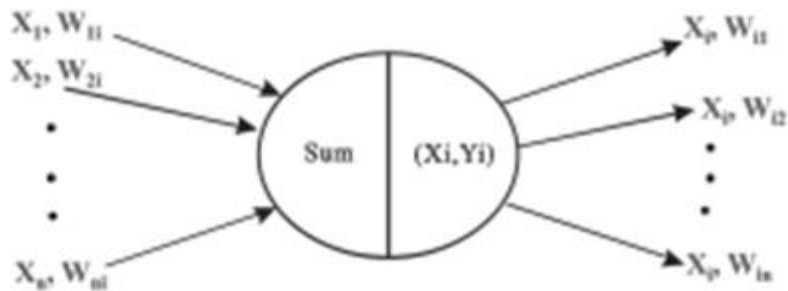

Oscillator to the polar coordinate system

$$\begin{bmatrix} \dot{r}_i \\ \dot{\theta}_i \end{bmatrix} = \begin{bmatrix} \frac{\tau_i(R_i - r_i)}{r_i} \\ \omega_i + r_i \sum_{j \neq i}^{n} w_{ij} r_j \sin(\theta_i - \theta_j - \phi_{ij}) \end{bmatrix}$$

where wij denotes the weight between Oscillator i and Oscillator j. wij = 0 means that there is no connection between Oscillators i and j.

Six oscillators are connected in a circle and are used to control the movement of the robotic fish. Neuron x in the oscillator is sent to drive the servo motor. In particular, generators 1 and 2 provide a control signal for the swing and swing motors of the left pectoral fin, while generators 3 and 4 are responsible for the swing and swing motors of the right pectoral fin, generators 5 and 6 control the left and right fin swing. tail fins, respectively. The PIC18F4550 microprocessor is responsible for determining the amplitude, frequency and phase difference between the various generators, as well as calculating the angular data for the motors according to the vibration equations.

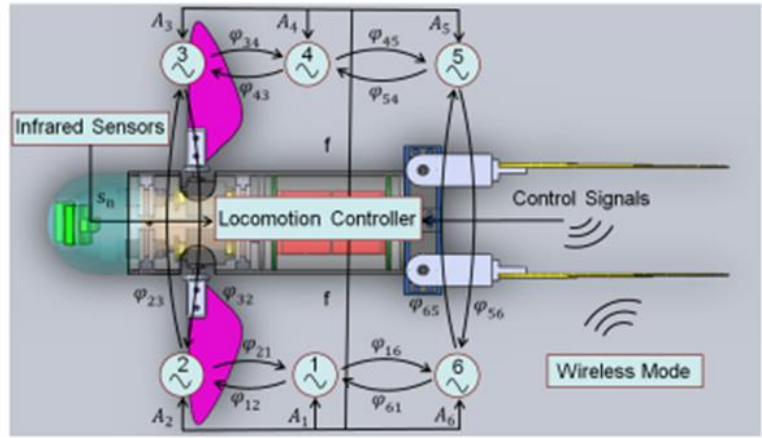

Fig. 8. Illustration of the CPG network utilized to control the robotic fish.

Underwater experiments on the robot have shown that the robot has high swimming characteristics and high maneuverability, which allows it to be used for research and performing tasks in a complex underwater environment. Autonomous swimming was also carried out with the feedback of three IR sensors. The results demonstrated the maneuverability and reliability of the proposed synthesized biomimetic robot. In addition, the robotic fish developed is very simple, cheap and modular, which also increases its potential applications.

**The main contribution of the article (Korkmaz, 2019)** is the development of a hierarchical control mechanism based on a CPG model with sensory feedback for performing three-dimensional multimodal swimming movements of robotic fish in an experimental pool in real time. Designed and developed biomimetic autonomous robotic robot with a two-link motor (i-RoF). As a basis for the hierarchical control mechanism, the CPG model based on the lamprey spinal cord from (Li, 2013) is used to provide a reliable and stable biomimetic control structure. The sensory neuron mechanism is adapted to the CPG model for the perception of external stimuli from the environment. In addition, the feedback fuzzy logic controller is designed as a brain model for identifying adaptive swimming patterns according to sensory information. A center of gravity (CoG) mechanism has been developed for up / down movements and can be controlled with a simple and convenient controller.

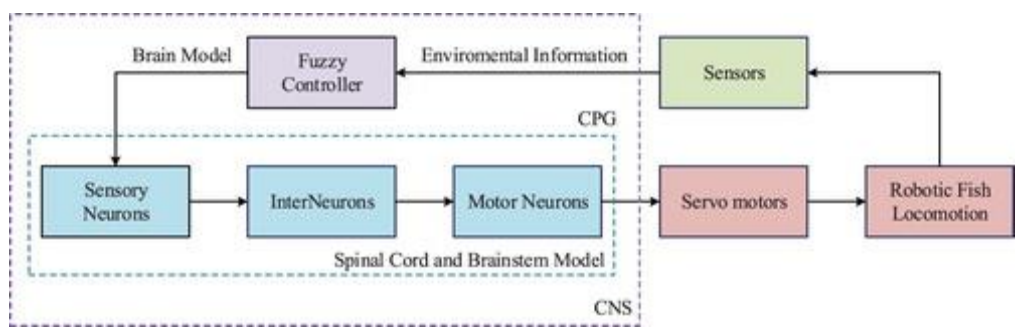

The CPG neural network consists of an engine control unit (MCU) and a series of chain oscillators that generate rhythmic oscillatory signals. The lamprey's neural oscillator is divided into two main parts, which are left and right symmetrical sections to each other. There are three interneurons in both sections, namely crossed interneurons (CIN), lateral interneurons (LIN), excitatory interneurons (EIN), and motor neuron (MN). Rhythmic oscillatory activity is provided by the interaction of these neurons. At the same time, each oscillator is linked by external communication synapses. The oscillator sends inhibitory synapses to other oscillators via EIN. Each LIN accepts incoming synapses from other generators. It was noted that the presynaptic neuron in the oscillator and the postsynaptic neuron in the other oscillator are on the same side as the left and right regions.

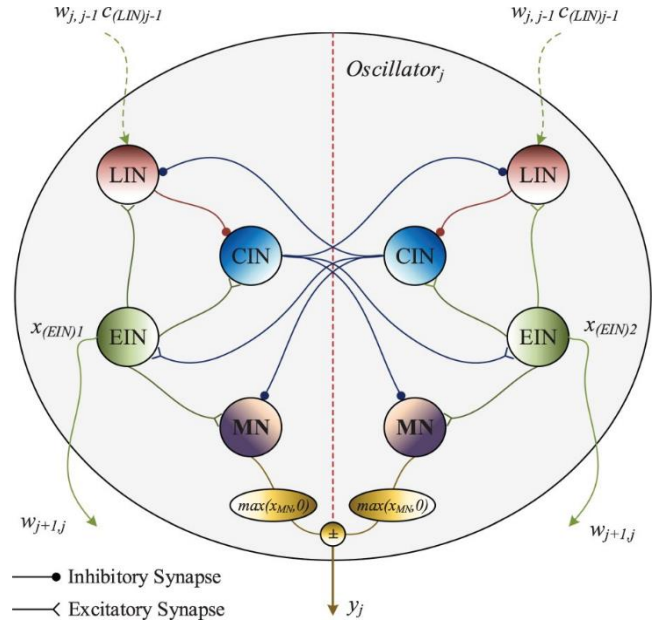

In the oscillator model, interneurons originate from a series of Leaky Integrate and Fire (LIF) cells. In the developed generator, each CIN can be defined as:

$$\tau \dot{x}_{(CIN)i} = -x_{(CIN)i} + \sum_{k=1}^{n} w_{i,k} s_{(CIN)i,k}$$
$$\{i = 1, 2; j = 1, 2, \ldots, N; k = 1, 2, \ldots, n\}$$

(1)

The LIN and EIN interneurons are expressed as Equations (2) and (3), respectively;

$$\tau \dot{x}_{(LIN)i} = -x_{(LIN)i} + \sum_{k=1}^{n} w_{i,k} s_{(LIN)i,k}$$
$$+ \sum_{j=1}^{N} w_{j,j-1} c_{(LIN)j-1}$$

(2)

$$x_{(EIN)i} = \frac{A}{1 + e^{x_{(\overline{CIN})}}} - \frac{A}{2}$$

(3)

The MN can be given by;

$$\tau \dot{x}_{(MN)i} = -x_{(MN)i} + \beta A + \sum_{k=1}^{n} w_{i,k} s_{(MN)i,k}$$

(4)

and thus, the oscillator output is obtained as below;

$$y_j = \max(x_{(MN)1}, 0) - \max(x_{(MN)2}, 0)$$

(5)

In these equations, i is the left and right sides belonging to the oscillators, j is the oscillator number, and k is the number of neurons in each oscillator. The upper line of the lower symbol defines the opposite side of the same oscillator. xi is the membrane potential of neurons, and yj is the output signal of the generator. τ, β and A define the period, offset and amplitude of the generator output, respectively. wik and wij are synaptic weights of neurons and synaptic weights of oscillators, respectively. si are the potentials of the presynaptic neuron in the same oscillator, and si with wik are the amount of neurotransmitter transferred to the postsynaptic neuron. cj is the presynaptic potential of a neuron in another oscillator, and cj with wij also gives the amount of neurotransmitter transmitted to the postsynaptic neuron in another oscillator. From equation (3) it can be seen that the neuron of the EIN type differs from the equation of the LIF neuron model. There is a sigmoid function with a threshold value A. EIN in the oscillator emits excitatory synapses to other neurons on the same side, and the sigmoid function provides constant changes in the membrane potential of neurons. The result is steady-state rhythmic oscillatory outputs.

The phase differences of the outputs in the model are determined by the MCU. The neural MCU acts as a brain stem in the spinal cord and has inhibitory synapses to oscillators according to a specific connection topology. The MCU is designed using two additional synapses (wR, L, wL, R) in the oscillator model. wR, L is the inhibitory synapse from the left EIN to the right LIN, and wL, R is the excitatory synapse from the right EIN to the left LIN. The phase difference between the oscillators is adjusted using these two additional synapses. To accurately adjust the phase difference, wR, L and wL, R must be in the range (0.1] and [-1.0), respectively, and they are defined by the following expression:

$$\begin{bmatrix} w_{L,R} & w_{R,L} \end{bmatrix} = \begin{bmatrix} \dfrac{\alpha}{\alpha+\gamma} & \dfrac{-\gamma}{\alpha+\gamma} \end{bmatrix}$$

For rhythmic and smooth sine generator outputs, all internal weights are set to 1, and the internal weights from EIN to MN are also set to 0.1. At the same time, the initial value of the neuron plays an important role in the generation of rhythmic oscillations. The output of a self-sustaining generator is obtained when the initial state is determined by a very small difference between the left and right CINs, which emit inhibitory synapses to neurons other than EIN.

**Sensory feedback mechanism**

The environmental data obtained by the sensors during the swimming of the robotic fish is estimated and the parameter λ, which is the input SN, is determined by the necessary fuzzy controller in accordance with the movement of the robotic fish determined by the decision mechanism. The interpreted stimulus received by the SN is sent to the CPG model, and the response produced by the robot fish and its output is sent to the servo motors via the MN. Thus, the swimming movements of the robot are obtained in accordance with the desired behavior. In the developed lamprey CPG oscillator with sensory feedback, each SN is determined by the following equation:

$$\begin{cases} \tau_{res}\dot{x}_{(SN)i} = -x_{(SN)i} + p\lambda & \Lambda \leq \lambda \\ \tau_{rec}\dot{x}_{(SN)i} = -x_{(SN)i} & \Lambda > \lambda \end{cases}$$

Here, x (SN) i is the membrane potential SN, τres is the rise time constant of the response generated to the stimulus, τrec is the decay time constant of the response generated to the stimulus, p is the stimulus correction factor. , λ - stimulus value, Λ - threshold value. The threshold determines when external stimuli become important and a response should be produced. When the magnitude of the stimulus is greater than or equal to the threshold value, the SN provides an output signal. Otherwise, the SN output is zero. In the proposed CPG model with sensory feedback, each SN emits excitatory synapses to all MNs. In experimental studies, for the SN parameters, τrec = 0.4, τres = 0.1095, p = 2, and Λ = 1. In addition, the values of the SN parameters in the left and right sections are equal to each other.

The specific features of the proposed CPG model sensory feedback system are as follows: (1) the proposed CPG provides a suitable solution for easily integrating a sensory feedback mechanism based on neurobiological structure into real-time applications; (2) the developed closed-loop control system behaves like an artificial model of the central nervous system and performs autonomous swimming against external disturbances; (3) sensory information is analyzed through the fuzzy controller and transmitted to the CPG model as sensory input; (4) up / down movements can be easily performed with the designed standard controller.

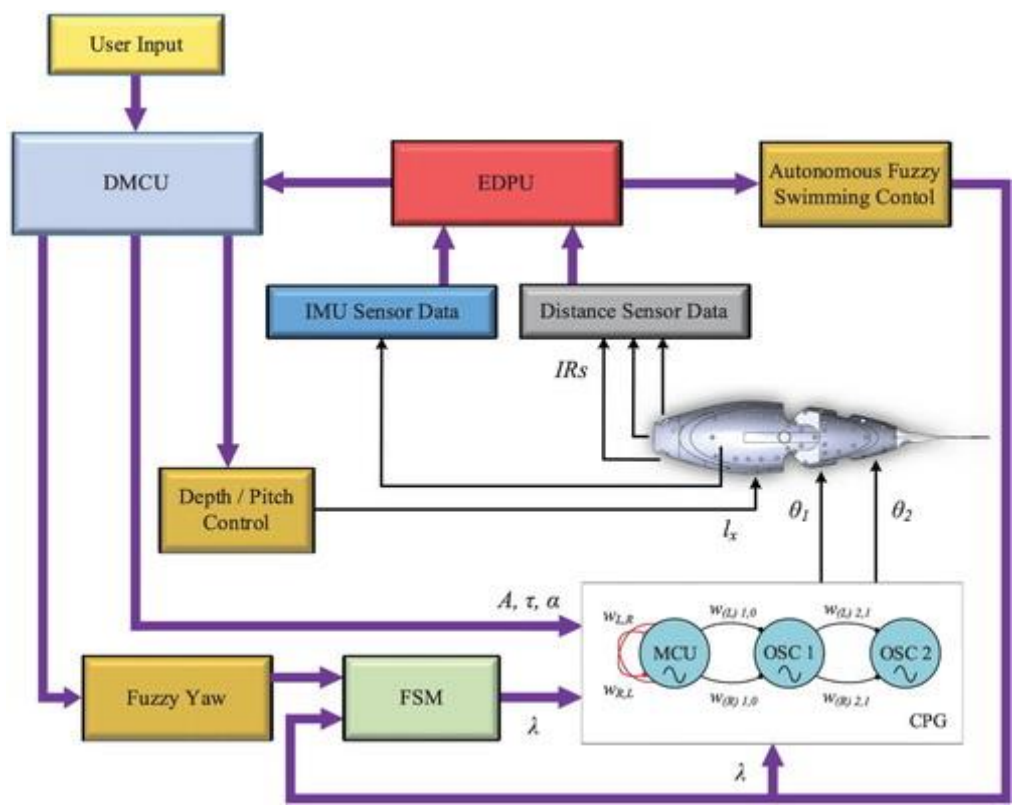

The experimental results contain three specific float modes such as yaw, pitch, and level hold, and these experiments are conducted with different scenarios. Yaw control is carried out according to three landmarks. Pitch control is shown for the two desired up / down movements, and finally level control near the base and near the surface is achieved. All results show that the proposed control structure provides an efficient and reliable response to actual fish movements.

## Salamander

### general description

The study of salamanders is a significant direction in the development of ideas about the evolutionary development of the CPG. This species is a transitional form of life, capable of rapidly changing various types of locomotor activity. In particular, the transition from rhythmic activity of the aquatic type to movements that provide ground movement seems to be interesting. Stimulation of the mesencephalic locomotor region both in the salamander and in all other classes of vertebrates causes movement, while the different strength of the stimulus determines the change in the type of movement from slow to fast. In the CFR of salamanders, two types of waves are very well traced: running and standing. Traveling waves are characteristic of walking motion on land and floating in water. Standing ones provide the ability to run quickly. Explanation: when running, the body bends; when standing, it is "immobilized".

**Description of the Knüsel Robot Salamander Model (Ijspeert, 2007)**

Non-linear coupled oscillators are used. The body's CPG uses 8 pairs of oscillators, with each pair responsible for its own node in the articulated body. It is important that when connected in series and in pairs, the oscillators have both direct and feedback. In this case, the feedback also goes from the CGH of the limbs to the CGH of the body. It is not clear what the 2 body oscillators are doing, which are opposite the limb oscillators (they do not have their own node).

Like lamprey models, burst (not sure what analogue in Russian) the properties of the vibrational center - fluctuations between bursts of motor neuron activity and rest periods - are modeled using a phase oscillator with a controlled amplitude (formulas are presented in the article).

The connections between the oscillators in the body's CPG are tuned to form traveling waves. The CPG of the extremities has lower frequencies of activity than the CPG of the body (accordingly, at high levels of arousal, standing waves are generated (the limb does not move)). The connections from the CGH of the limbs to the CGH of the body are stronger than in the opposite direction. CGH of the extremities, when activated, transfers the entire CGH into a walking state.

In this case, the motors in the robot are controlled based on the difference between the bursts in paired oscillators ("from the left and right bodies")

**Complicated**

An ideological follow-up was a study published by Harischandra in 2011 (Harischandra). Scientists have investigated the influence of sensory communication on the work of the CGH of the salamander. However, it is noteworthy that the CGR consisted of 800 (500E and 300I) LIF neurons. In this case, the structure contained 40 axial segments, consisting of identical parts (right and left; flexor and extensor), including excitatory, inhibitory, and moto subnets. In this case, one part, upon activation, inhibits the second. E neurons connected 1 segment rostrally and 3 caudally, and I neurons 2 segments rostral and 6 caudally. It is necessary to clarify that for the sake of simplicity, the authors limited themselves to 14 pairs of body generators. Moreover, each limb was a complex CGH of 3 pairs of complex parts (their composition is described above), due to which the limb had three degrees of freedom in accordance with the anatomical structure.

Interestingly, in this study, the researchers reported a more successful trot gait with strengthening connections from the nodes of the limbs to the nodes of the body.

Abstract: the CGH scheme is generally identical to the study scheme (1), however, instead of oscillators, we used simulators consisting of subpopulations of LIF neurons

**The next step in the development** of ideas about the salamander's CGR was the introduction of various kinds of asymmetric connections between the rostracaudal segments into the models, as well as differences in the structure of individual CGH nodes. This approach, presented in (3), demonstrates an explanation of the change in the types of movement and is valid for CGH on oscillators and LIF neurons.

In the CGR, intersegmental connections were directed only caudally. Axial network of 16 segments. In contrast to previous works, the CPG of the extremities is represented by simple networks and only generates a rhythm (a simplified limb is obtained, the authors admit that more complex CPG is required for the limbs). In addition, inhibitory connections from limb CPG go only to the two proximal segments of the CPG body with a decreasing likelihood of joining.

The various states of the CRG operation are assessed by two parameters: the frequency of oscillations or spikes and the phase delay or spikes in successive segments.

The authors propose using the concept of multistable neural networks. They demonstrate that the described approach allows one network not only to store within itself various generated patterns in the form of a phase delay, but also to reproduce them after various types of external influences on individual parts of the generator.

On oscillators, this is achieved by limiting the propagation of rostral and caudal influences (in the CGH, connections are removed from the main part of the CGH of the body to the first and last segments): the effect of establishing a uniform intersegment phase lag along the cord without limiting the actual value of the phase lag.

The connections between the CGH segments of the body were also changed. In the rostro-caudal direction, they were five times stronger than in the opposite direction. Thanks to this, the body's CPG was divided into fast front and slow back parts. Accordingly, the phase delays were now adjusted by the initial phase lag of the segments. In addition, now the choice of gait or swimming was carried out due to the influence of the CGR of the limbs on the anterior oscillators. These conclusions are based on the CFR of the lamprey from the work of Kozlov 2009.

**Also, one cannot** fail to mention the work Bicanski, A., 2013, in which, as well as in the above-described study, the importance of interscillatory connections for the most successful variant of the functioning of the CPG was emphasized. However, the work also had a serious drawback: there were no motoneurons in the CGR networks either. A simplified version of XX was used here.

**Subsequently, the attention** of the scientific community was attracted by works that focused on greater biological relevance. In 2018, Liu, Wang proposed a type of CGR called Locomotion-controlled neural networks (Liu, Q., & Wang, J. Z., 2018).

LCNNS consists of a new neuron model that reflects the spike character of neurons in the lamprey spine. These new Lcnns can describe most of the properties of real biological neural networks and can be used to build salamander neural networks based on the burst generator model proposed by Guertin (2009).

**In the same year,** the authors of this model, as part of the team of authors (Liu, Q., Yang, H., Zhang, J., & Wang, J. (2018)), published a paper directly on the modeling of salamander CGH on spike neurons. It postulates the need for different networks for each type of movement and speed, as well as a complex connection scheme between the networks, both within the CGH of the extremities and inside the CGH of the body. In this case, the low-frequency CGH of the extremities is associated only with the low-frequency parts of the CGH of the body. This is supported by the latest data obtained (McLean, D., Masino, M, 2008).

In the model, all neurons of the high-frequency part of the global CGH have the same frequency as in the low-frequency one. Moreover, the networks are also the same in topological terms. Those. the robot is controlled by a global CGR, which consists of two separate networks of the same size, but with different operating frequencies (the model scheme is at least twice as large as in previous works, since there are more networks). The same goes for the CPG of the extremities. At the same time, descending excitatory connections predominate in the body's CPG. The minimum unit of the CGR is a network of two neurons with mutual inhibition. It should be noted that in such networks, the delays between segments do not lend themselves to separate regulation; accordingly, the entire network has the same spike frequency as an individual neuron. In this case, the frequency response does not depend on the amplitude.

**A direct continuation of these** studies is (Liu, Q., Zhang, Y., Wang, J., Yang, H., & Hong, L. (2020)), in which the authors modify their model in order to achieve greater biological plausibility. The new version eliminates a serious drawback: now the model contains not only interneurons, but also motoneurons. In addition, the connections between the segments of both the body CGH and the limb CGH were modified. As a result, the authors demonstrated the ability to control the direction of the first movement, the ability to make a U-turn. In addition, a sensory connection in the form of stretch receptors has been added to the model.

**In the article by** Knüsel, J., Crespi, A., Cabelguen, J.-M., Ijspeert, A. J., & Ryczko, D. (2020), the authors continue to study the issue of the influence of sensory connections on the work of CGR. Unlike his previous research, here the robot is controlled by a network of oscillators of 25 pairs of segments. The robot's tail is a passive fin.

The new model proved to be capable of reproducing signals similar to recordings from isolated preparations. In particular, the researchers even succeeded in showing spontaneous switching of CGR from slow caudorostral to fast rostracaudal waves (with a change in the exciting impulse).

This study is also interesting because it demonstrates the absence of the need for the presence of several CPGs to create patterns of different properties, as previously assumed /

**Another study that postulates** the need for sensory feedback is the work (Suzuki, S., Kano, T., Ijspeert, A. J., & Ishiguro, A. (2021)). Here, as in the previous study, the authors demonstrate a spontaneous transition between different patterns. However, the CPG model uses different feedback rules: feedback from limb to limb, limb to body, body to limb, and body to body without any connection between oscillators. The first rule is responsible for the coordination of the four legs as they move forward, supporting the body. The second and third rules involve cross-feedback, which establishes self-organized body-limb coordination. The fourth rule coordinates the lateral irregularities of the multi-segment trunk.

Those. in the model, a flexible change in the CGH parameters is possible due to the point effect on each individual oscillator with a signal from the "brain" (analogous to tonic stimulation).

This is the first study to demonstrate the spontaneous transition of gait from horizontal walking with standing body waves to walking trot with traveling waves.

**This work was supported by the Russian Science Foundation project № 21-12-00246**